\documentclass[10pt,conference]{IEEEtran} %ICSE
\IEEEoverridecommandlockouts
\usepackage{graphicx}
\usepackage{textcomp}

\usepackage{soul}
\usepackage{multirow}
\usepackage{hyperref}

\usepackage{CJKutf8}

\usepackage[edges]{forest}
\definecolor{hidden-draw}{RGB}{20,68,106}
\definecolor{hidden-pink}{RGB}{255,245,247}
\definecolor{mydraw}{RGB}{150, 150, 150}
\definecolor{mypink}{RGB}{255, 182, 193}
\usepackage{enumitem}
\usepackage{multicol}
\usepackage{colortbl}  %彩色表格需要加载的宏包
\usepackage{xcolor}
\usepackage{amsmath,amsfonts}
\usepackage{algorithm}
\usepackage{algpseudocode}
\usepackage{ulem}
\usepackage[switch]{lineno}
\usepackage{listings, xcolor}
\definecolor{verylightgray}{rgb}{.97,.97,.97}
\lstdefinelanguage{Solidity}{
  keywords=[1]{anonymous, assembly, assert, balance, break, call, callcode, case, catch, class, constant, continue, constructor, contract, debugger, default, delegatecall, delete, do, else, emit, event, experimental, export, external, false, finally, for, function, gas, if, implements, import, in, indexed, instanceof, interface, internal, is, length, library, log0, log1, log2, log3, log4, memory, modifier, new, payable, pragma, private, protected, public, pure, push, require, return, returns, revert, selfdestruct, send, solidity, storage, struct, suicide, super, switch, then, this, throw, transfer, true, try, typeof, using, value, view, while, with, addmod, ecrecover, keccak256, mulmod, ripemd160, sha256, sha3}, % generic keywords including crypto operations
  keywordstyle=[1]\color{blue}\bfseries,
  keywords=[2]{address, bool, byte, bytes, bytes1, bytes2, bytes3, bytes4, bytes5, bytes6, bytes7, bytes8, bytes9, bytes10, bytes11, bytes12, bytes13, bytes14, bytes15, bytes16, bytes17, bytes18, bytes19, bytes20, bytes21, bytes22, bytes23, bytes24, bytes25, bytes26, bytes27, bytes28, bytes29, bytes30, bytes31, bytes32, enum, int, int8, int16, int24, int32, int40, int48, int56, int64, int72, int80, int88, int96, int104, int112, int120, int128, int136, int144, int152, int160, int168, int176, int184, int192, int200, int208, int216, int224, int232, int240, int248, int256, mapping, string, uint, uint8, uint16, uint24, uint32, uint40, uint48, uint56, uint64, uint72, uint80, uint88, uint96, uint104, uint112, uint120, uint128, uint136, uint144, uint152, uint160, uint168, uint176, uint184, uint192, uint200, uint208, uint216, uint224, uint232, uint240, uint248, uint256, var, void, ether, finney, szabo, wei, days, hours, minutes, seconds, weeks, years},  % types; money and time units
  keywordstyle=[2]\color{teal}\bfseries,
  keywords=[3]{block, blockhash, coinbase, difficulty, gaslimit, number, timestamp, msg, data, gas, sender, sig, value, now, tx, gasprice, origin},  % environment variables
  keywordstyle=[3]\color{violet}\bfseries,
  identifierstyle=\color{black},
  sensitive=false,
  comment=[l]{//},
  morecomment=[s]{/*}{*/},
  commentstyle=\color{gray}\ttfamily,
  stringstyle=\color{red}\ttfamily,
  morestring=[b]',
  morestring=[b]"
}
\usepackage{pifont}
\usepackage{diagbox}
\usepackage{subfigure}
\usepackage{tcolorbox}
\usepackage{color}
\usepackage{microtype}

\usepackage{oplotsymbl}
\usepackage{subfigure}
\usepackage{siunitx}
\usepackage{array,framed}
 \usepackage{
   color,
   float,
   epsfig,
   wrapfig,
   graphics,
   graphicx,
 }
\usepackage{setspace}
\usepackage{amsfonts}
\usepackage{latexsym,fancyhdr,url}
\usepackage{enumerate}
\usepackage{algpseudocode}
\usepackage{graphics}
\usepackage{xparse} % argument parsing -- \edist
\usepackage{xspace}
\usepackage{multirow}
\usepackage{microtype}
\usepackage{soul}
\soulregister\cite7 
\soulregister\ref7 

\usepackage{color}
\usepackage[T1]{fontenc}
\usepackage{fancyhdr}

\usepackage{subfigure}
\usepackage{booktabs}
\usepackage{cite}
\usepackage{balance}

\usepackage{CJKutf8} %%显示中文的包

\newcommand{\todo}[1]{}

\newcommand{\myauthornote}[3]{}

\newcommand{\chong}[1]{\myauthornote{Chong}{blue}{#1}}

\newcommand{\later}[1]{}
\newcommand{\boxmargin}{1mm}
\tcbuselibrary{most, skins, breakable, theorems}
\newtcolorbox{myboxa}[2][]{
    colback=gray!10!white,
    colframe=black, enhanced,
    attach boxed title to top left={yshift=-2mm,xshift=5mm},
    title=#2,#1
}
\newtcolorbox{myboxb}[2][]{
    boxsep=1pt,
    left = \boxmargin, right = \boxmargin, top = \boxmargin, bottom = \boxmargin,
    title={#2},#1
}
\newtcolorbox{myboxc}{
    colback=gray!15!white,
    arc = 0pt, outer arc = 0pt,
    boxsep=0pt, left = 3pt, right = 0pt, top = 0pt, bottom = 0pt, 
    leftrule=3pt, bottomrule=0pt,toprule=0pt, rightrule=0pt,
    left = \boxmargin, right = \boxmargin, top = \boxmargin, bottom = \boxmargin
}
\newcommand{\figmargin}{\vspace{-2mm}} %-4mm
\newcommand{\tabmargin}{\vspace{0mm}} %-3mm
\begin{document}
\begin{CJK}{UTF8}{gkai}

\title{LLM Hallucinations in Practical Code Generation: Phenomena, Mechanism, and Mitigation}
% \maketitle

\author{
\IEEEauthorblockN{Ziyao Zhang$^{1}$, Yanlin Wang$^{1}$\IEEEauthorrefmark{1}, Chong Wang$^{2}$, Jiachi Chen$^{1}$, Zibin Zheng$^{1}$}
\IEEEauthorblockA{$^{1}$Sun Yat-sen University, Zhuhai, China}
\IEEEauthorblockA{$^{2}$Nanyang Technological University, Singapore}
\thanks{\IEEEauthorrefmark{1} Corresponding author.}
}
\maketitle

\begin{abstract}
Code generation aims to automatically generate code from input requirements, significantly enhancing development efficiency. Recent large language models (LLMs) based approaches have shown promising results and revolutionized code generation task. Despite the promising performance, LLMs often generate contents with hallucinations, especially for the code generation scenario requiring the handling of complex contextual dependencies in practical development process. Although previous study has analyzed hallucinations in LLM-powered code generation, the study is limited to standalone function generation. In this paper, we conduct an empirical study to study the phenomena, mechanism, and mitigation of LLM hallucinations within more practical and complex development contexts in repository-level generation scenario. First, we manually examine the code generation results from six mainstream LLMs to establish a hallucination taxonomy of LLM-generated code. Next, we elaborate on the phenomenon of hallucinations, analyze their distribution across different models. We then analyze causes of hallucinations and identify four potential factors contributing to hallucinations. Finally, we propose an RAG-based mitigation method, which demonstrates consistent effectiveness in all studied LLMs. The replication package including code, data, and experimental results is available at \url{https://github.com/DeepSoftwareAnalytics/LLMCodingHallucination}.
\end{abstract}

% \maketitle
%    <concept>
%        <concept_id>10011007.10011074.10011099</concept_id>
%        <concept_desc>Software and its engineering~Software verification and validation</concept_desc>
%        <concept_significance>500</concept_significance>
%        </concept>
%  </ccs2012>

% \ccsdesc[500]{Software and its engineering~Software verification and validation}

% \ccsdesc[500]{Software and its engineering~Software verification and validation}

% \keywords{Code generation, Hallucination, Large language models}




\section{Introduction}
Code generation is an automation technology aimed at efficiently producing code from specifications described in natural language. This process significantly reduces the manual coding workload for developers~\cite{chen2021evaluating, barke2023grounded, zhang2023unifying}, allowing them to focus more on solving advanced technical challenges and engaging in innovative tasks. Recent developments have introduced a variety of large language models (LLMs)~\cite{gpt-neo,gpt-j,Codex,CodeT5,CodeT5Plus,InCoder,AlphaCode,CodeGen,GPT-NeoX,SantaCoder,StarCoder,CodeLlama,zeng2021pangu,guo2024deepseek} built upon the Transformer architecture~\cite{Transformer}. These models, trained on extensive code corpora, can automatically generate code from natural language inputs and have shown high efficacy in code generation. For example, GPT-4 has achieved state-of-the-art results on evaluation benchmarks such as HumanEval~\cite{Codex} and MBPP~\cite{corr/abs-2108-07732}, demonstrating high functional correctness, 
particularly in generating \textit{standalone} functions based on detailed specifications.

However, in practical development scenarios, the requirements for code generation are more complex than simply generating standalone functions from detailed specifications~\cite{yu2024codereval}. To address this complexity, new benchmarks, such as CoderEval~\cite{yu2024codereval}, ClassEval~\cite{du2023classeval}, and EvoCodeBench~\cite{li2024evocodebench}, have been proposed to better reflect \textbf{\textit{real-world repository-level}} development scenarios. Evaluations based on these benchmarks have revealed that LLMs face challenges in generating \textit{non-standalone} functions with contextual dependencies, such as calls to user-defined functions and project-defined data protocol. While these benchmarks provide valuable insights into the effectiveness of LLMs in practical code generation, they primarily focus on functional correctness as measured by test case pass rates and lack a thorough analysis of underlying failure causes. \textit{To bridge this gap, this work aims to systematically investigate issues in practical LLM-based code generation from the perspective of hallucinations.}

Hallucination is a significant issue for state-of-the-art generative LLMs~\cite{huang2023survey}. For general natural language tasks, LLM hallucinations have been explored to a certain extent~\cite{tonmoy2024comprehensive,zhang2023siren,huang2023survey,ye2023cognitive} and are typically categorized into three types: Input-Conflicting Hallucination, Fact-Conflicting Hallucination, and Context-Conflicting Hallucination~\cite{zhang2023siren}. In the domain of code generation, Liu et al.~\cite{liu2024exploring} conducted a study to analyze hallucinations in LLM-powered code generation and established a taxonomy that aligns with these three categories. While Liu et al.'s study provided insightful findings, it is based on benchmarks (i.e., HumanEval~\cite{Codex} and DS-1000~\cite{lai2023ds}) for \textit{standalone} function/script generation instead. Our work, however, focuses on hallucinations within more practical and complex development contexts in repository-level generation scenarios. Additionally, their study primarily categorized hallucinations from a problem-presentation perspective to uncover fine-grained code-semantic issues, resulting in categories such as \textit{Dead Code} and \textit{Repetition}. In this work, we investigate hallucinations from a holistic perspective in terms of \textbf{\textit{phenomena}}, \textbf{\textit{mechanism}}, and \textbf{\textit{mitigation}}. We believe that our study can complement the findings by Liu et al., providing a broader understanding of hallucinations in LLM-based code generation.

In this work, we conduct an empirical study to uncover the status quo and root causes of hallucinations in LLM-based code generation within real-world projects. The study aims at answering the following research questions (RQs):
\begin{itemize}[leftmargin=15pt]
\item \textbf{RQ1 (Hallucination Taxonomy):} What are the specific manifestations of hallucinations in practical code generation, and how are they distributed?
\item \textbf{RQ2 (LLM Comparison):} How do different LLMs compare in terms of hallucination occurrences and patterns?
\item \textbf{RQ3 (Root Causes):} What are the root causes of hallucinations in practical LLM-based code generation?
\end{itemize}

To answer the questions, we experiment on six mainstream LLMs (ChatGPT~\cite{chatgpt}, CodeGen~\cite{CodeGen}, PanGu-$\alpha$~\cite{zeng2021pangu}, StarCoder2~\cite{lozhkov2024starcoder}, DeepSeekCoder~\cite{guo2024deepseek}, and CodeLlama~\cite{CodeLlama}) with the CoderEval dataset~\cite{yu2024codereval}. 
To obtain the hallucination taxonomy of practical LLM-based code generation, we manually perform open coding~\cite{khandkar2009open} on the LLM-generated code. 
Specifically, we first extract 10\% of the coding tasks from the CoderEval dataset in the initial stage. Then, from the initial annotation and discussion, we obtain preliminary taxonomy. Finally, we obtain the fully hallucination taxonomy with iterative labelling the remaining 90\% coding tasks and continuously refining the taxonomy in the process. After obtaining the taxonomy, we conduct extensive analysis based on the research questions aforementioned. 

\textbf{Findings.} 
Our study reveals the following findings. \chong{Show the findings...} 
\textcircled{1} LLM hallucinations in code generation can be divided into three major categories (Task Requirement Conflicts, Factual Knowledge Conflicts, and Project Context Conflicts) with eight subcategories: Functional Requirement Violation, Non-Functional Requirement Violation, Background Knowledge Conflicts, Library knowledge Conflicts, API Knowledge Conflicts, Environment Conflicts, Dependency Conflicts, and Non-code Resource Conflicts. 
\textcircled{2} We analyze the hallucination distribution in different LLMs and find that Task Requirement Conflicts are the most prevalent type of hallucination across all models. 
\textcircled{3} We identify four potential factors that cause hallucinations: training data quality, intention understanding capacity, knowledge acquisition capacity, and repository-level context awareness. 

\textbf{Mitigation.} 
Based on the findings, we explore a lightweight mitigation approach based on retrieval augmented generation (RAG) and evaluate its effectiveness.
In this approach, we construct a retrieval library based on the repository in the development scenario of each generation task and obtain the code snippet that is beneficial to the current generation task as a prompt through the similarity detection between the task description in the generation task and the code snippet in the retrieval library. Experimental results show that this lightweight mitigation can consistently improve the performance of all studied LLMs. 

In summary, this paper makes the following contributions:
\begin{itemize}[leftmargin=15pt]
    \item We conduct an empirical study to analyze the types hallucinations in LLM code generation in real development scenarios and establish a hallucination taxonomy in LLM-based code generation.
    \item We elaborate on the phenomenon of hallucinations, analyze the distribution of hallucinations on different models. %, and discuss the reasons for such situations further.
    \item We further analyze causes of hallucinations and identify four possible factors. 
    \item We propose a RAG-based mitigation approach based on the causes of hallucinations and experiment on various LLMs to study its effectiveness.
    \item We make the replication package available at \url{https://github.com/DeepSoftwareAnalytics/LLMCodingHallucination}, to support further studies in this field.
\end{itemize}

\section{Background \& Related Work}
\subsection{LLM-based Code Generation}
For developers, a realistic scenario is to use a code repository to write code, which is very common in practice~\cite{grechanik2010empirical}. For example, due to security and functionality considerations, companies often only build code warehouses internally. The code repository provides many private APIs that are not seen by the language model and are not public on any code hosting platform. Therefore, it is worth exploring whether pre-trained language models can adapt to real development needs and generate correct and efficient code. In real-world development scenarios, the development of a function not only relies on the text description and function signature of the function, but also requires calling a custom API in the code repository. Such non-independent functions are commonly found in real-world generation scenarios. By analyzing the 100 most popular projects written in Java and Python on GitHub~\cite{yu2024codereval}, previous work found that dependent functions account for more than 70\% of the functions in open source projects. In order to better simulate real development scenarios and to check the correctness of LLMs, CoderEval~\cite{yu2024codereval}, ClassEval~\cite{du2023classeval}, and EvoCodeBench~\cite{li2024evocodebench} collected code snippets and text descriptions from real code repositories and used test cases to check the correctness of the code repositories in their corresponding environments.However, the performance of the model on these benchmarks is extremely poor. LLMs cannot generate correct code based on the problem description, and the model prefers to generate independent code segments rather than using existing functions in the current development scenario.

\subsection{Hallucinations in LLMs}
In the field of natural language processing (NLP), hallucination refers specifically to situations where the content produced by a language model in the process of generating text is inconsistent with the given input or expected output environment, lacks meaning, or violates the facts~\cite{ji2023survey}. This kind of phenomenon is particularly prominent in text generation models, especially in tasks such as text completion, summary generation, and machine translation. The output of the model must maintain a high degree of consistency and authenticity to ensure its practicality and reliability.
Hallucination phenomena can be divided into the following categories according to their nature~\cite{zhang2023siren}: (1) Input-Conflicting Hallucinations: When the text generated by the model deviates from the original input source, input-conflicting hallucinations will occur. This illusion may result from the model's incorrect parsing or inaccurate internal representation of the input information, causing the output content to deviate from the intent and context of the source input. (2) Context-Conflicting Hallucinations: This type of hallucination occurs when the text generated by the model is contradictory or inconsistent with its previously generated content. Contextual conflict hallucinations reflect the model's challenges in maintaining textual coherence and consistency, which may be due to the model's insufficient processing of contextual information or limitations of its memory mechanism. (3) Fact-Conflicting Hallucinations: When the content generated by LLM is inconsistent with established knowledge or facts in the real world, fact-conflicting hallucinations will occur. This illusion reveals the model's inadequacy in understanding and applying knowledge about the external world, and may be caused by limitations in model training data, lags in knowledge updates, or limitations in the model's reasoning capabilities.

However, there is a lack of research on hallucination phenomena in the field of code generation. Although there have been a large number of LLM-based methods to optimize code generation tasks, these works do not have a clear definition of the code generation illusion. The presence of hallucination problems can be detrimental to the overall quality of the generated code. This may not only affect the performance and maintainability of the code, but may also lead to unexpected errors and security vulnerabilities, thus posing a threat to the stability and security of the software. In order to make up for the gaps in the definition of hallucination problems, there has been work to define hallucinations for LLMs in code generation tasks. This work~\cite{tonmoy2024comprehensive} defined new hallucination standards for LLMs in code generation tasks and divided hallucinations into five main types, but this work ignores that LLMs in real-world code generation tasks will involve relevant knowledge unique to the software engineering field such as development environment, system resources, external constraints, code warehouses, etc. These factors often cause LLMs to fail in actual development. Problems such as low usability and low accuracy. In order to better explore the illusions that exist in LLMs in real development scenarios, our work obtained data sets in real development scenarios for empirical study, and defined new types of illusions, which opened up new ideas for subsequent research on illusions.

\section{Evaluation Setup}\label{sec:setup}
% \subsection{Task Collection}
\subsection{Dataset}
To better simulate practical development scenarios, we use a set of coding tasks from real-world Python repositories based on the CoderEval benchmark~\cite{yu2024codereval}. CoderEval comprises 230 Python code generation tasks, extracted from a diverse set of Python repositories. Each task consists of a natural language description, a ground-truth code snippet, and a set of test cases, along with the project environment context associated with the task.

% \subsection{LLM-based Generation}
\subsection{Studied LLMs}
We utilize several mainstream LLMs to perform code generation for the studied programming tasks. The LLMs being used cover both open-source and closed-source models and span various parameter sizes, listed as follows. %\chong{briefly introduce the models}

\begin{itemize}[leftmargin=15pt]
\item \textbf{ChatGPT}~\cite{chatgpt}: ChatGPT is a versatile text generation model for multilingualism with powerful code generation capabilities, we use the GPT-3.5-Turbo in our experiments.

\item \textbf{CodeGen}~\cite{nijkamp2022codegen}: CodeGen is a family of auto-regressive language models for program synthesis with  several different versions. To better accomplish the generation task, we use the  CodeGen-350M-Mono model.

\item \textbf{PanGu-$\alpha$}~\cite{zeng2021pangu}: PanGu-$\alpha$ can perform code generation tasks in multiple languages. We use the PanGu-$\alpha$-2.6B model.

\item \textbf{DeepSeekCoder}~\cite{guo2024deepseek}: DeepSeekCoder performs well in open source models across multiple programming languages and various benchmarks. We use the DeepSeekCoder-6.7B  base model.

\item \textbf{CodeLlama}~\cite{CodeLlama}: CodeLlama is a set of pre-trained and fine-tuned generative text models ranging in size from 7 to 34 billion parameters. We use the CodeLlama-7b-Python-hf model.

\item \textbf{StarCoder2}~\cite{StarCoder}: StarCoder2 is a family of open code-oriented models for large languages, providing three scales of models, we use the StarCoder2-7B model.
    
\end{itemize}

For each task, we use the LLMs to generate 10 code snippets by employing  
the nuclear sampling strategy and setting temperature to 0.6, following the same setting as CoderEval. 

% \subsection{Manual Annotation}
\subsection{Taxonomy Annotation}
\label{sec:taxonomy_annotation}
In order to analyze the hallucination types in the LLM-generated code, we manually perform open coding~\cite{khandkar2009open} on the generated code to obtain the hallucination taxonomy. 

\textbf{(1) Initial Open Coding.} Firstly, in the initial open-coding stage, we select 10\% of the 230 coding tasks in CoderEval Python dataset for preliminary analysis. We randomly collect 23 generative tasks from CoderEval, we employ CodeGen, Pangu-$ \alpha $, ChatGPT, DeepSeekCoder, CodeLlama, and StarCoder2, with each model generating ten code snippets for each code generation task, culminating in a total of 1,380 code snippets to be analysed for hallucination taxonomy framework. For each code snippet, we test it in the actual development environment corresponding to the task to determine its correctness. On this basis, the two authors will compare the differences between the ground-truth and LLMs generated code snippets and discuss and record possible hallucination phenomena. 

\textbf{(2) Preliminary Taxonomy Construction.} Secondly, we document possible hallucinations in the generated code and the location of the hallucination content. Several different hallucinations may occur within a single code snippet. 
All annotators are required to discuss the codes and define the code's hallucinatory taxonomy. In this process, we classify similar hallucination to create a preliminary taxonomy that illustrates the various hallucination types and their meanings in the code generated by LLMs. 

\textbf{(3) Full Taxonomy Construction.} Finally, after obtaining the categorisation criteria, the remaining code snippets will be independently annotated by three newly invited volunteers with extensive Python programming experience, two with more than ten years of experience and one with four years of programming experience. If new types of hallucinations arise that are not covered by the current taxonomy, annotators are required to write descriptions of the hallucinations to allow further discussion to establish new types and enhance the taxonomy.

%%%%========Taxonom BEGIN===========
\tikzstyle{my-box}=[
    rectangle,
    draw=hidden-draw,
    rounded corners,
    text opacity=1,
    minimum height=1.5em,
    minimum width=5em,
    inner sep=2pt,
    align=center,
    fill opacity=.5,
    line width=0.8pt,
]
\tikzstyle{leaf}=[my-box, minimum height=1.5em,
    fill=hidden-pink!80, text=black, align=left,font=\normalsize,
    inner xsep=2pt,
    inner ysep=4pt,
    line width=0.8pt,
]
\begin{figure*}[t]
\centering
\resizebox{0.99\textwidth}{!}{
\begin{forest}
    forked edges,
    for tree={
        grow=east,
        reversed=true,
        anchor=base west,
        parent anchor=east,
        child anchor=west,
        base=center,
        font=\large,
        rectangle,
        draw=hidden-draw,
        rounded corners,
        align=left,
        text centered,
        minimum width=4em,
        edge+={darkgray, line width=1pt},
        s sep=3pt,
        inner xsep=2pt,
        inner ysep=3pt,
        line width=0.8pt,
        ver/.style={rotate=90, child anchor=north, parent anchor=south, anchor=center},
    },
    where level=1{text width=19em,font=\normalsize,}{},
    where level=2{text width=19em,font=\normalsize,}{},
    where level=3{text width=22em,font=\normalsize,fill=hidden-pink!40}{},
    % where level=4{text width=7em,font=\normalsize,}{},
    [
        Code Generation Hallucination, ver
        [
            Task Requirement Conflicts (\S\ref{sec:Requirement-Conflicts})
            [
                Functional Requirement Violation %(\S\ref{sec:functional-requirement-mismatch})
                    [{Example: Wrong Functionality,  Missing Functionality}]
            ]
            [
                Non-functional Requirement Violation %(\S\ref{sec:non-functional-requirement-violation})
                    [{Example: Security, Performance, Style, Code Smell}]
            ]
        ]
        [
            Factual Knowledge Conflicts (\S\ref{sec:Factual-Knowledge-Conflicts})
            [
                Background Knowledge Conflicts %(\S\ref{sec:Background-Knowledge-Conflicts})
            ]
            [
                Library Knowledge Conflicts %(\S\ref{sec:Library-Knowledge-Conflicts})
            ]
            [
                API Knowledge Conflicts %(\S\ref{sec:API-Knowledge-Conflicts})
                    [{Example: Parameter, Guard Conditions \\ Similar-but-wrong APIs, Exception Handling}]
            ]
        ]
        [
            Project Context Conflicts (\S\ref{sec:Project-Context-Conflicts})
            [
                Environment Conflicts %(\S\ref{sec:Development-Environment-Conflicts})
            ]
            [
                Dependency Conflicts %(\S\ref{sec:Dependency-Conflicts})
                    [{Example: Undefined Methods, API Version Conflict}]
            ]
            [
                Non-code Resource Conflicts %(\S\ref{sec:Non-code-Resource-Conflicts})
                    [{Example: Data, Config, Assert, Connection}]
            ]
        ]
    ]
\end{forest}
}
\figmargin
\caption{Taxonomy of Hallucinations in LLM-based Code Generation}
\label{fig:taxonomy}
\end{figure*}
%%%%========Taxonomy END===========

\section{Evaluation Results}
In this section, we present the evaluation results and answer the three aforementioned research questions.

\subsection{RQ1: Hallucination Taxonomy}
\begin{figure}[t]
    \centering
    \setlength{\abovecaptionskip}{0.1cm}\includegraphics[width=0.9\linewidth]{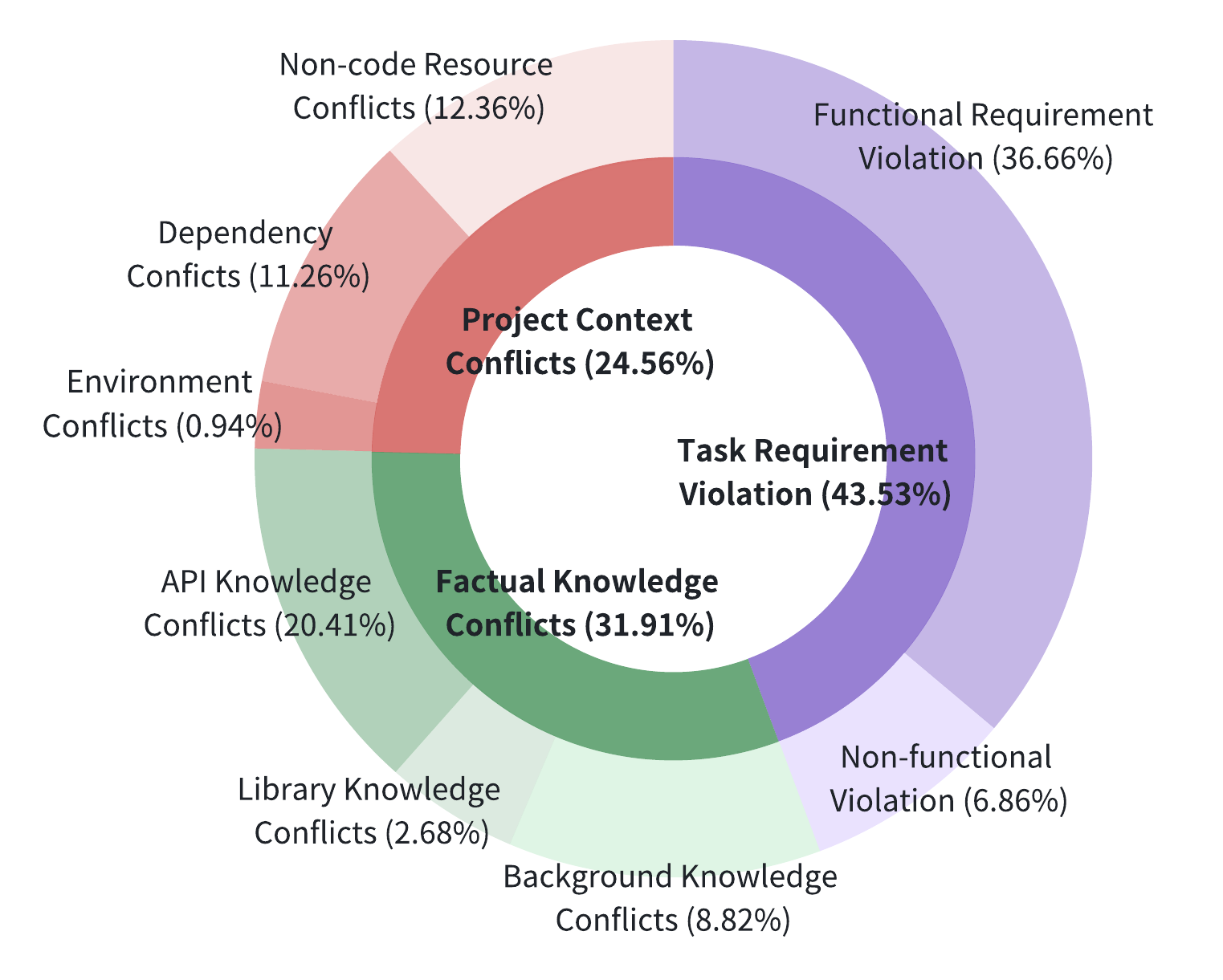}
    \figmargin
    \caption{Hallucination Distribution}
    \label{fig:Distribution-of-hallucinations}
\end{figure}

The overall LLM coding hallucination taxonomy we obtained from Section~\ref{sec:taxonomy_annotation} is presented in Figure~\ref{fig:taxonomy}. Through manual annotation, we identify three primary hallucination categories: Task Requirement Conflicts, Factual Knowledge Conflicts, and Project Context Conflicts, which can be further divided into eight specific types. Note that our three primary categories align well with the hallucination types in the general domain~\cite{ji2023survey}. 
% \begin{itemize}[leftmargin=15pt]
{Task requirement conflicts} correspond to \textit{input-conflicting hallucinations} in the general domain, indicating that the generated code does not meet the functional or non-functional requirements of the coding tasks.
{Factual knowledge conflicts} correspond to \textit{knowledge-conflicting hallucinations} in the general domain, indicating that the generated code does not comply with background knowledge, library/framework knowledge, or API knowledge.
{Project context conflicts} correspond to \textit{context-conflicting hallucinations} in the general domain, indicating that the generated code incorrectly uses project contexts, including environments, dependencies, and resources.
% \end{itemize}
In the following, we present the detailed hallucination types in our taxonomy. Figure~\ref{fig:Distribution-of-hallucinations} shows the distribution of the hallucination types.

\subsubsection{\textbf{Task Requirement Conflicts (43.53\%)}}
\label{sec:Requirement-Conflicts}
In the general domain, input-conflicting hallucinations occur when the answers generated by LLMs deviate from the original intentions of user inputs~\cite{huang2023survey}. In the context of code generation tasks, the primary intentions of inputs typically revolve around the functional and non-functional requirements of the coding tasks. When the code generated by LLMs does not align with these requirements, hallucinations related to Task Requirement Conflicts occur. Specifically, these conflicts can be categorized into two types: {Functional Requirement Violation} and {Non-functional Requirement Violation}.

\begin{figure}[t]
    \centering
    \includegraphics[width=\linewidth]{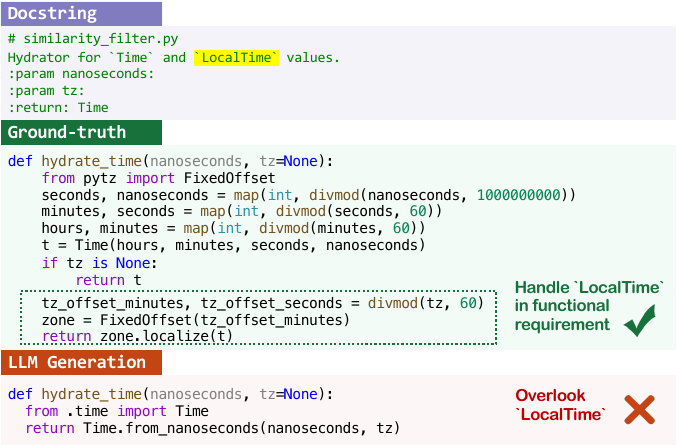}
    \figmargin
    \figmargin
    \caption{Example: Functional Requirement Violation}
    \label{fig:functional-requirement-voilation}
\end{figure}

\underline{Functional Requirement Violation (36.66\%).}
\label{sec:functional-requirement-mismatch}
Functional requirements are typically expressed in natural language and describe the desired functionality of the generated code. When these requirements are not correctly and comprehensively understood, the resulting code may fail to meet expected functionality, leading to logic bugs (such as unexpected execution behaviors) or runtime errors (such as the \texttt{KeyError} during dictionary access). More specifically, the functional requirement mismatch can be subdivided into two typical types: \textit{Wrong Functionality} and \textit{Missing Functionality}. 
    For example, as illustrated in Figure~\ref{fig:functional-requirement-voilation}, the functional requirement involves handling \texttt{LocalTime} based on the specific timezone \texttt{tz}. In the ground-truth code, this requirement is addressed by the lines highlighted in the green rectangle. However, the code generated by PanGu-$\alpha$ overlooks this requirement, resulting in a hallucination of Functional Requirement Violation.

\begin{figure}[t]
    \centering
    \setlength{\abovecaptionskip}{0.1cm}
    \includegraphics[width=\linewidth]{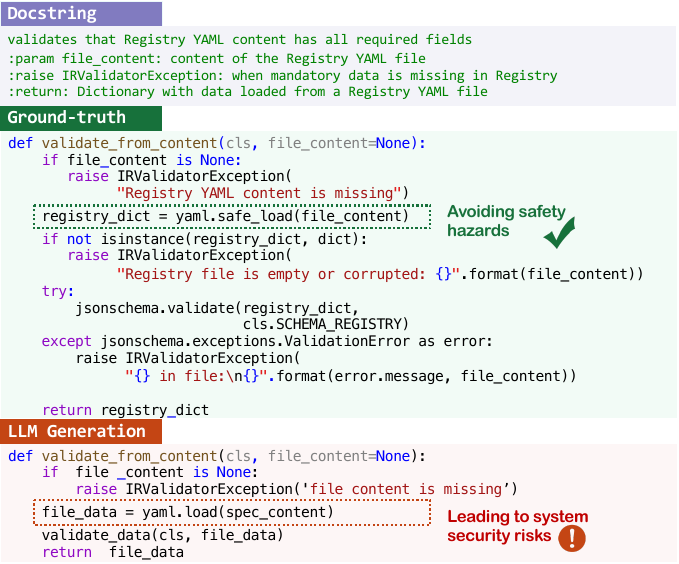}
    \figmargin
    \figmargin
    \caption{Example: Non-functional Requirement Violation}
    \label{fig:Non-functional-Requirement-Violation}
\end{figure}

\underline{Non-functional Requirement Violation (6.86\%).}
\label{sec:non-functional-requirement-violation}
Besides functional requirements, developers often have non-functional requirements for the generated code, such as security concerns or performance considerations. These non-functional requirements are usually more implicit than functional requirements and are not described in the input natural language descriptions. Our open coding annotation reveals that non-functional requirements in coding tasks can be mainly divided into the following aspects: \textit{Security}, \textit{Performance}, \textit{Style}, and \textit{Code Smell}. Generated code that violates these non-functional requirements may introduce safety risks or increase the maintenance complexity of the corresponding project.

Specifically, on the security side, the generated code may introduce vulnerabilities such as unsanitized inputs, which can lead to insecure deserialization or SQL injection attacks. As shown in Figure~\ref{fig:Non-functional-Requirement-Violation}, the ground-truth code uses the \texttt{safe\_load} function to safely read YAML files. In contrast, the LLM-generated code utilizes the \texttt{load} function, thereby introducing a potential security risk.
Regarding performance, the generated code may lack optimization for execution efficiency, for example, by using inefficient loop structures that lead to unnecessary overhead in computing and memory resources. 
Style violations often occur when the generated code fails to follow established programming conventions or style guides, such as inconsistent naming conventions or inappropriate code layout, which can negatively affect code readability and maintainability.
Code smell violations include issues such as overly complex functions or excessive use of global variables, which increase the complexity and potential risks associated with future maintenance.

\subsubsection{\textbf{Factual Knowledge Conflicts (31.91\%)}}
\label{sec:Factual-Knowledge-Conflicts}
In the field of NLP, the term ``factual conflicts'' refers to content generated by LLMs that does not align with established knowledge or facts about the real world. Practical software development similarly relies on various types and levels of factual knowledge to produce correct code. Consequently, when LLMs fail to accurately understand and apply background knowledge~\cite{wang2023beyond}, library/framework knowledge, or API knowledge, hallucinations on Factual Knowledge Conflicts arise. 
We further divide this hallucination category into three types:  {Background Knowledge Conflicts}, {Library Knowledge Conflicts}, and {API Knowledge Conflicts}.
    
\underline{Background Knowledge Conflicts (8.82\%).}
\label{sec:Background-Knowledge-Conflicts}
Background Knowledge Conflicts are a common issue when using large language models. These conflicts refer to the situation that the generated code is inconsistent with existing domain-specific knowledge, potentially rendering the code invalid or introducing logic bugs and risks. For instance, in automotive software development, if the generated code fails to adhere to certain industry standards (e.g., AUTOSAR\footnote{https://en.wikipedia.org/wiki/AUTOSAR}), it can result in significant compliance issues or safety risks.

\begin{figure}[t]
    \centering
    \setlength{\abovecaptionskip}{0.1cm}
    \includegraphics[width=\linewidth]{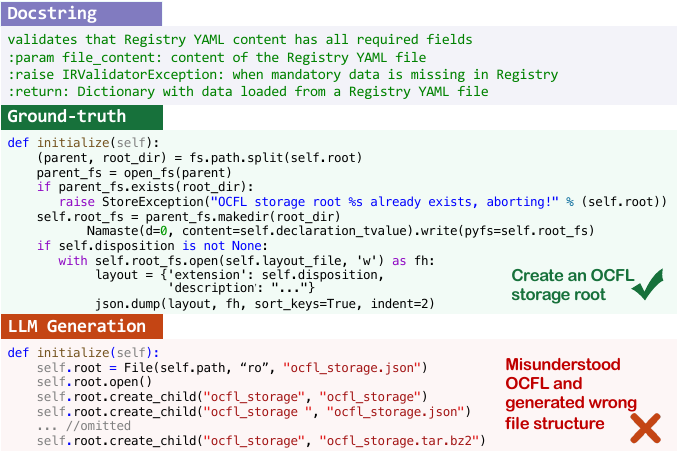}
    \figmargin
    \figmargin
    \caption{Example: Background Knowledge Conflicts}
    \label{fig:Background-Knowledge-Conflicts}
\end{figure}

Background knowledge typically includes \textit{Domain Concepts} (e.g., specific data formats or protocols) and related \textit{Standards and Specifications} (e.g., standard parameters or configurations). For example,  Figure~\ref{fig:Background-Knowledge-Conflicts} shows an example about OCFL (Oxford Common File Layout), a specification for data storage and transformation. According to the official description\footnote{https://ocfl.io/1.1/spec/\#storage-root}, an OCFL storage root must contain a ``Root Conformance Declaration'' following the ``NAMASTE'' specification and may include a file named \texttt{ocfl\_layout.json} to describe the root layout arrangement. While the ground-truth code aligns with these specifications when initializing the OCFL storage root, the generated code might incorrectly focus on other OCFL aspects that are irrelevant to creating the storage root. 

\begin{figure}[t]
    \centering
    \setlength{\abovecaptionskip}{0.1cm}
    \includegraphics[width=\linewidth]{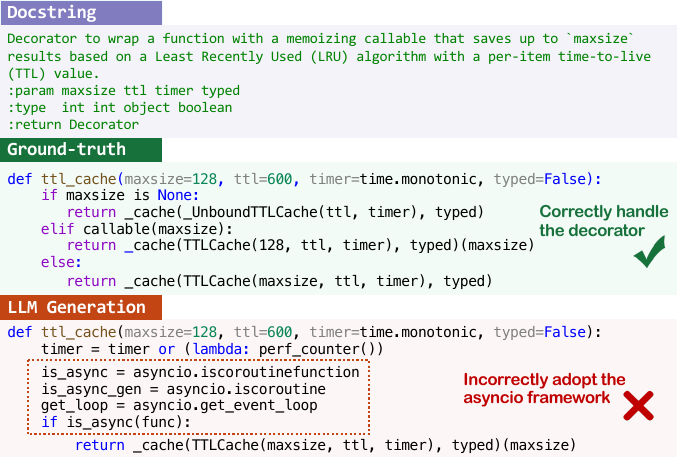}
    \figmargin
    \figmargin
    \caption{Example: Library Knowledge Conflicts}
    \label{fig:Library-Knowledge-Conflicts}
\end{figure}

\underline{Library Knowledge Conflicts (2.68\%).}
\label{sec:Library-Knowledge-Conflicts}
In modern software development, developers frequently employ frameworks or third-party libraries (e.g., Django\footnote{https://www.djangoproject.com/} for web applications) to expedite the development process by reusing the features or functionalities that these frameworks or libraries provide. When utilizing these frameworks or libraries, LLMs may encounter factual errors that lead to unexpected behaviors or even security risks. For example, As depicted in Figure~\ref{fig:Library-Knowledge-Conflicts}, the task requires the model to generate a decorator that caches the return value of the function upon each invocation. In the code generated by the DeepSeekCoder model, the APIs from the \texttt{asyncio} framework are utilized. This framework is designed for asynchronous processing, and the model's misuse of the asynchronous processing framework poses unexpected behaviors to the developed application.

\underline{API Knowledge Conflicts (20.41\%).}
\label{sec:API-Knowledge-Conflicts}
API Knowledge Conflicts are a common hallucination in LLM-generated code caused by various types of API misuses, such as parameter errors, improper guard conditions, similar-but-incorrect/deprecated API usage, and improper exception handling. 
For example, parameter errors can occur when inappropriate parameter types or values are used in the generated code, causing API calls to fail or return unexpected results. This case is especially common in dynamically typed programming language such as Python~\cite{wang2024tiger}. Improper guard conditions mean that the generated code does not correctly implement pre-condition checks. If the validity of the pre-conditions of certain APIs is not verified before calling them (e.g., file existence), runtime errors may occur. In terms of similar-but-wrong/deprecated API usage, LLMs may mistakenly choose APIs with similar functions but different applicable scenarios. Although this choice is syntactically correct, it cannot meet actual application needs. Improper exception handling involves generating code that fails to properly handle potential exceptions, which can cause the program to crash or behave abnormally when faced with an error condition. This kind of API knowledge conflict will not only directly lead to program functional errors, but may also affect the stability of the system and the usability of the code. 

\begin{figure}[t]
    \centering
    \includegraphics[width=\linewidth]{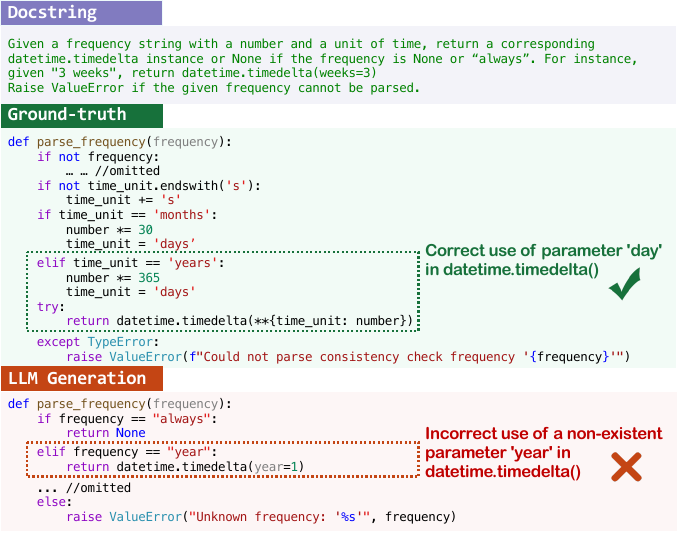}
    \figmargin
    \figmargin
    \caption{Example: API Knowledge Conflicts}
    \label{fig:API-Knowledge-Conflicts}
\end{figure}

We present an example in Figure~\ref{fig:API-Knowledge-Conflicts}. In this generation task, CodeGen correctly identifies the task intent and utilizes the \texttt{datetime.timedelta()} function. However, the code snippet generated by CodeGen uses a non-existing parameter \texttt{year}.

% \end{itemize}

\subsubsection{\textbf{Project Context Conflicts (24.56\%)}}
\label{sec:Project-Context-Conflicts}
Project Context Conflict hallucination refers to the phenomenon where the code generated by LLMs is inconsistent with the specific context of a given project. In a sense, this type of hallucination is also a type of factual conflict, where facts within the current project context are violated. The key difference is that Factual Knowledge Conflicts involve common facts (e.g., libraries and APIs) that are publicly accessible, while Project Context Conflicts pertain to facts that are specific to the corresponding project, which are generally unavailable for public access. Project Context Conflicts are often caused by LLMs not aware of such project-specific facts when generating code. This hallucination can be divided into \textit{Environment Conflicts}, \textit{Dependency Conflicts}, and \textit{Non-code Resource Conflicts}.

\begin{figure}[t]
    \centering
    \includegraphics[width=\linewidth]{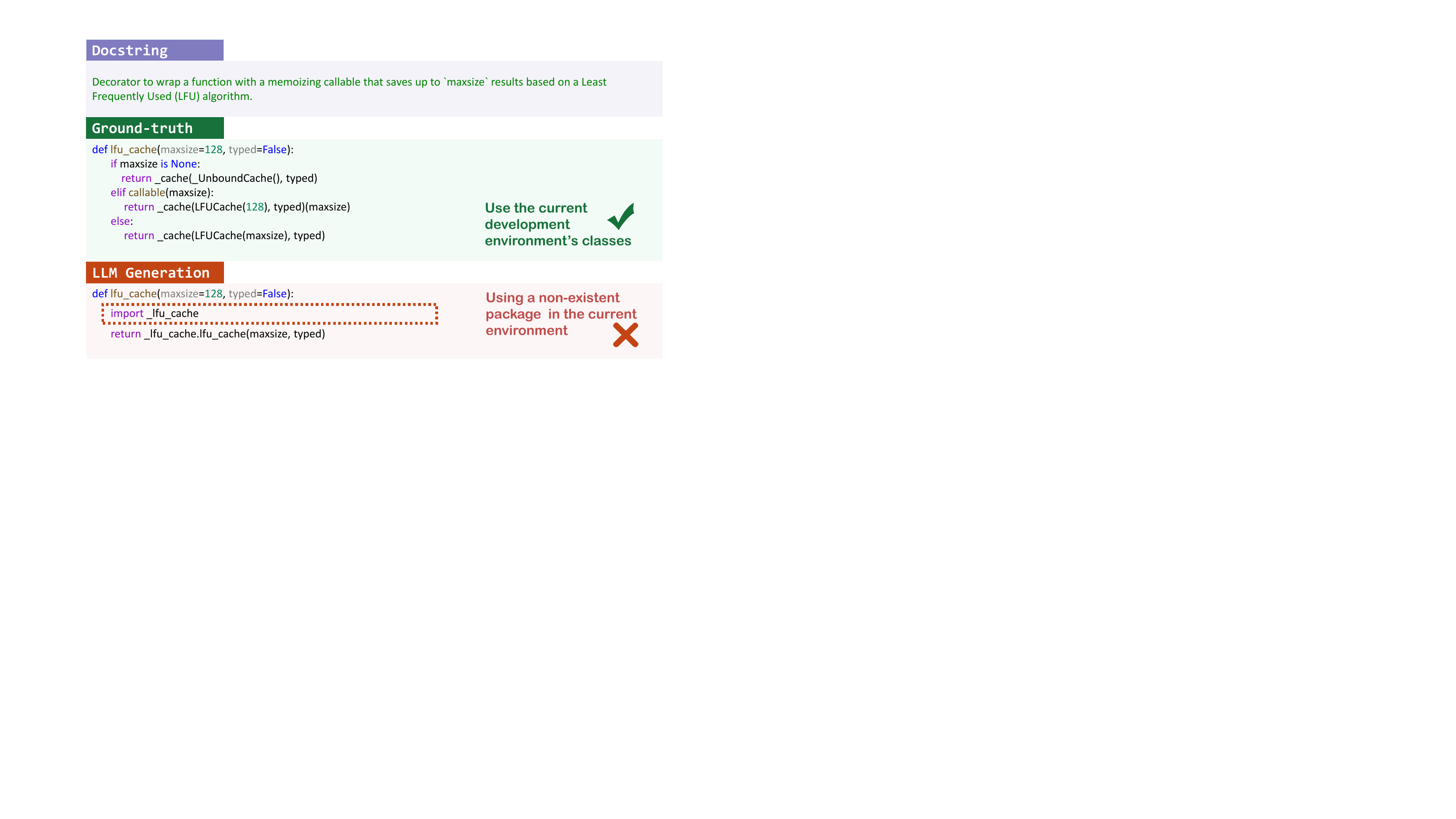}
    \figmargin
    \figmargin
    \caption{Example: Environment Conflicts}
    \label{fig:Development-Environment-Conflict}
\end{figure}

% \begin{itemize}[leftmargin=15pt]
\underline{Environment Conflicts (0.94\%).}
\label{sec:Development-Environment-Conflicts}
In the process of software development, conflicts between the generated code and the development environment are common, especially regarding version differences in platforms, operating systems, drivers, languages, compilers/interpreters, frameworks, and libraries. When generating code, such environmental concerns are often not considered, leading to problematic code if there are environment-sensitive operations. For example, if the generated code uses language features (e.g., \texttt{f}-string expressions) from higher Python versions that are not supported by the current development environment, a conflict arises. 
For example, Figure~\ref{fig:Development-Environment-Conflict} shows a code snippet generated by CodeGen that attempts to use the package \texttt{\_lfu\_cache}, which does not exist in the current environment.

\underline{Dependency Conflicts (11.26\%).}
\label{sec:Dependency-Conflicts}
Dependency Conflicts arise when the generated code relies on undefined or unimported dependencies, such as user-defined attributes and functions. This often results in errors such as undefined variables or no-member errors. In practical software development, 70\% of functions are non-standalone and depend on entities defined elsewhere in the project or imported from third-party libraries~\cite{yu2024codereval}. Due to the inability of LLMs to access the entire project context, they often resort to using non-existent APIs, functions, attributes, and variables when dealing with non-standalone functions. 

\begin{figure}[t]
    \centering
    % \captionsetup{labelfont=bf, name=Example, labelsep=period}
    \setlength{\abovecaptionskip}{0.1cm}
    \includegraphics[width=\linewidth]{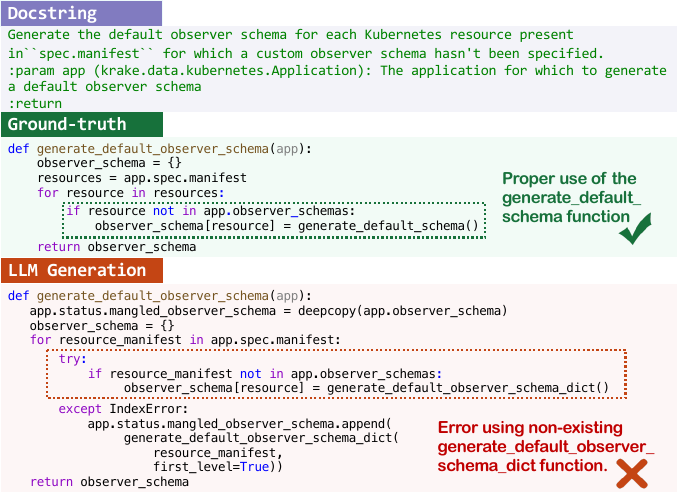}
    \figmargin
    \figmargin
    \caption{Example: Dependency Conflicts}
    \label{fig:Dependency-Conflicts}
\end{figure}

For example, Figure~\ref{fig:Dependency-Conflicts} illustrates a scenario involving a user-defined function \texttt{generate\_default\_observer\_schema\_dict()}. In this case, the PanGu-$ \alpha $ erroneously uses a function with a similar but incorrect name, \texttt{generate\_default\_schema()}, which does not exist in the project. This leads to a Dependency Conflict, as the code fails to execute correctly due to the missing definition.
    
\underline{Non-code Resource Conflicts (12.36\%).}
\label{sec:Non-code-Resource-Conflicts}
    Non-code Resource Conflicts can be further categorized into four main types: \textit{Data}, \textit{Configs}, \textit{Assets}, and \textit{Connections}. Each type of conflict can undermine the correctness and reliability of the system.
    Data conflicts often involve mishandling of data formats, fields, or content. For example, if the generated code incorrectly parses a data file or attempts to access a non-existent data field, it can lead to runtime errors or data inconsistencies.
    Config conflicts arise from incorrect settings or options in configuration files. This might include using undefined configuration fields or options, which can prevent the generated code from properly applying the configuration and affect system behavior.
    Asset conflicts here involve improper handling of asset files and their properties. For instance, if the generated code fails to set the correct size and resolution for images or videos, it can result in display issues or severe bugs, such as application crashes.
    Connection conflicts relate to wrong settings of various connection resources, such as incorrect IP addresses, port numbers, or database tables. These issues often lead to failed connections or operations being performed on the wrong server or database, potentially causing data leaks or security incidents. 
    
\begin{figure}[t]
    \centering
    % \captionsetup{labelfont=bf, name=Example, labelsep=period}
    \setlength{\abovecaptionskip}{0.1cm}\includegraphics[width=\linewidth]{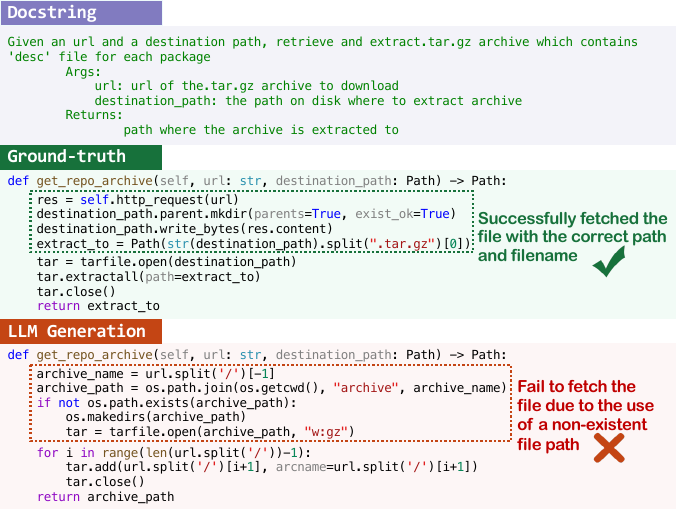}
    \figmargin
    \figmargin
    \caption{Example: Non-code Resource Conflicts}
    \label{fig:Resource-Misuses}
\end{figure}

    For example, Figure~\ref{fig:Resource-Misuses} illustrates the generation task hopes that LLM can generate a function for a given URL and target path to retrieve and extract the \texttt{tar.gz} compressed package containing each package's ``description'' file. However, in the code snippet generated by the model, the model adds ``archive'' as a path in the target path, which causes the code snippet to point to a non-existent file path. This will not allow the \texttt{tar.gz} compressed package to be correctly obtained, resulting in a program error.

% \input{src/cause}

% \end{itemize}

\begin{center}
     
    \begin{myboxc}{\textbf{RQ1 Summary: }
    We have established a hallucination taxonomy in LLM-based code generation, comprising  three main categories (i.e., Task Requirement Conflicts, Factual Knowledge Conflicts, and Project Context Conflicts) with eight subtypes. Among these, Task Requirement Conflicts are the most frequently occurring category.
    }
    \end{myboxc}
    
\end{center}

\subsection{RQ2: LLM Comparison}
\begin{figure}[t]
    \centering
    % \captionsetup{labelfont=bf, name=Example, labelsep=period}
    % \setlength{\abovecaptionskip}{0.1cm}
    \includegraphics[width=0.95\linewidth]{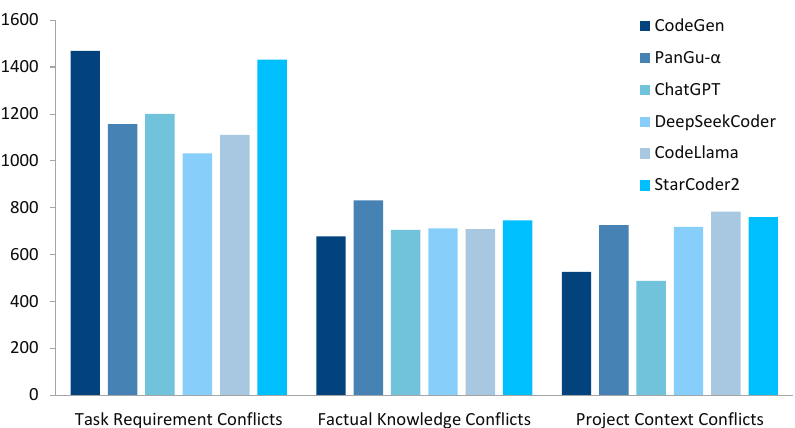}
    \figmargin
    \caption{Hallucination distribution of different models}
    % \Description{A PDF image showing Distribution of Different models}
    \label{fig:Distribution-of-models}
\end{figure}

Based on the obtained hallucination taxonomy for LLM-based code generation, we further analyze the hallucination distribution comparison across different models.      
Figure~\ref{fig:Distribution-of-models} shows the distribution of the number of hallucinations of different models based on the breakdown analysis of the three hallucination types. 
We find that Task Requirement Conflicts are the most common hallucination type for all models, while Factual Knowledge Conflicts and Project Context Conflicts remain at approximately the same frequency. 
Additionally, we find that CodeGen and StarCoder2 exhibit a notably higher frequency of hallucinations related to Task Requirement Conflicts, whereas DeepSeekCoder and CodeLlama demonstrates the lowest occurrence. This variation may be related to the models' ability to understand task requirements, potentially influenced by factors such as the model size or the training corpora. For instance, DeepSeekCoder and CodeLlama are trained on diverse corpora including both extensive code and text data, while CodeGen and StarCoder2 are primarily trained on code-related data.
In terms of factual knowledge conflicts, PanGu-$\alpha$ demonstrates the highest frequency of factual hallucinations. This can be attributed to its extensive training on Chinese corpora, which may have led to a relatively limited exposure to factual knowledge, such as specific domain concepts, expressed in English.

\begin{center}
     
    \begin{myboxc}{\textbf{RQ2 Summary: }
    Task Requirement Conflicts are the most prevalent type of hallucination across all models, with CodeGen and StarCoder2 showing a notably higher frequency of this type compared to others. 
    }
    \end{myboxc}
    
\end{center}

\subsection{RQ3: Root Cause Analysis}
In this research question, we conduct further analysis on the possible root causes of the hallucinations in practical LLM-based code generation.
 
\subsubsection{\textbf{Training Data Quality}}
The quality of the training data is a crucial factor in the development of LLMs, as it significantly affects models' inference capabilities. Recent LLMs are often trained on large-scale code corpora typically collected from open-source repositories. However, the quality of these repositories is not always assured, leading to the inclusion of low-quality data in the training corpora. Such issues include mismatches between docstrings and code~\cite{sun2022importance}, inefficient or insecure code implementations~\cite{klemmer2024using}, misused API calls, outdated library documentation and usage~\cite{zhong2024can}, and a lack of domain diversity. 
When LLMs are trained on such corpora, they may unintentionally incorporate these flaws into their knowledge base, leading to hallucinations in code generation. As shown in Figure~\ref{fig:Non-functional-Requirement-Violation}, LLMs may generate code that uses unsafe APIs, reflecting problematic patterns commonly found in the training data. 
This indicates that the model may have been affected by low-quality data during the training phase. 
Most hallucinations associated with Task Requirement Conflicts and Factual Knowledge Conflicts can be, to a certain extent, attributed to data quality issues in the training corpora. This highlights the importance of building a high-quality code-related training data to reduce hallucinations in code generation.

\subsubsection{\textbf{Intention Understanding Capacity}}
Although LLMs have shown great potential in code generation, they still face challenges in accurately capturing and interpreting specific user intentions and needs~\cite{mu2023clarifygpt}. This limitation can result in generated code that is functionally or non-functionally inaccurate, thereby affecting the overall effectiveness and trustworthiness of LLM-based code generation~\cite{spiess2024quality}. The core advantage of LLMs lies in their excellent pattern recognition capabilities, but this is also the source of their limitations. LLMs tend to generate code based on common patterns observed in the training data rather than from a deep understanding of the specific requirements context.
As shown in Figure~\ref{fig:functional-requirement-voilation}, the task description requires the LLM to handle \texttt{LocalTime}, which is ignored in the LLM-generated code. This example highlights LLMs' inadequacy in comprehensively interpreting the  intentions behind requirements. Furthermore, LLMs also show limitations in handling subtle requirements involving complex logic or multi-step operations~\cite{zan2022language}. Due to a poor understanding of the overall scope and potential limitations of the task, LLM-generated code may only address part of the requirements or perform poorly in handling edge cases. This can result in generated code snippets that seem correct on the surface but fail to meet specific business logic or functional requirements in practice.

\subsubsection{\textbf{Knowledge Acquisition Capacity}}
LLMs may learn incorrect knowledge and miss certain domain-specific knowledge due to the aforementioned training data quality issues. Moreover, as software development techniques evolve, such as library updates, relevant knowledge developed after model training period cannot be acquired by LLMs. Unlike human developers who can continuously learn and integrate latest information during development, LLMs are limited to the knowledge available at the time of training. 
For example, as shown in Figure~\ref{fig:Background-Knowledge-Conflicts}, the task description needs a piece of code for generating a data format that satisfies the OCFL storage specification, but LLM generates incorrect code, possibly due to its lack of  the OCFL-related knowledge during inference.
This limitation in LLMs' knowledge acquisition capacity leads to hallucinations related to incorrect or outdated factual information in the generated code. This highlights the need for a knowledge acquisition mechanism, such as retrieval augmented generation (RAG), to allow LLMs to update, correct, and supplement the knowledge they have learned.

\subsubsection{\textbf{Repository-level Context Awareness}}

Feeding all project contexts, including code, documents, and non-code resources, into an LLM for repository-level code generation is challenging and impractical. This is because LLMs, typically based on the Transformer architecture~\cite{Transformer}, have token number limits (e.g. 8k or 12k tokens) and experience quadratic computation growth as the number of tokens increases. 
Additionally, including all project contexts can introduce a significant amount of irrelevant information, hindering LLMs' ability to focus on the most relevant context for code generation. Therefore, it is crucial to develop methods that make LLMs aware of the project contexts (project-specific memory) that are precisely related to the current coding task. Recent works attempt to integrate static analysis tools~\cite{wang2024teaching} or apply retrieval-augmented generation (RAG) based on repository-level retrieval corpora~\cite{zhang2023repocoder} to address such context awareness issues.

\begin{center}
    % \begin{myboxb}[]{RQ4 Summary} 
    \begin{myboxc}{\textbf{RQ3 Summary: }
    By further analyzing the causes of hallucinations, we identify four possible contributing factors: training data quality, intention understanding capacity, knowledge acquisition capacity, and repository-level context awareness. Deficiencies in any of these factors can lead to hallucinations in practical development scenarios. 
    }
    \end{myboxc}
    % \end{myboxb}
\end{center}

\section{Mitigation Approach}
\label{sec:mitgation_approach}
\subsection{Motivation}
The aforementioned root causes of hallucinations in code generated by LLMs can be traced back to three main factors at the inference stage: incorrect or insufficient understanding for task requirements, the lack of factual knowledge pertinent to the generation tasks, and the inability to access the necessary code and non-code resources from the repository. 
These limitations create substantial challenges for LLMs in code generation in practical development settings. There are many previous works investing LLM-based code generation~\cite{wang2024rlcoder,li2024repomincoder,guo2024stop,wang2024beyond,zheng2023towards,zheng2023survey,zheng2023towards,wang2021code,nie2023unveiling}, we draw inspiration from existing work~\cite{zhang2023repocoder} on repository-level code generation and explore the feasibility of applying retrieval-augmented generation (RAG) to mitigate hallucinations.The idea is that by providing LLMs with code snippets relevant to the current task, they can better understand the requirements and gain awareness of specific factual knowledge and project contexts.

\subsection{RAG-based Mitigation}
To implement the RAG method, we first collect all code repositories from the CoderEval dataset and follow RepoCoder's method~\cite{zhang2023repocoder} to construct the retrieval corpora. Specifically, for each repository, we apply a sliding window to scan all the source files in it. This scanning process extracts consecutive lines of code based on a predefined window size. The sliding window moves by a fixed number of lines (slicing step) at each iteration to ensure complete coverage of the code. We adhere to RepoCoder's parameter settings, with a window size of 20 lines and a sliding step of 2 lines. To prevent answer leakage, code lines containing or following the ground-truth code are excluded from the scanning process. Once all files are processed, a retrieval corpus of code snippets is generated for the repository.

We employ a sparse bag-of-words (BOW) model for our retrieval mechanism, which simplifies gauging similarity between textual data. This model transmutes both the query and the candidate code snippets into sets of tokens, which are compared using the Jaccard index. The Jaccard index measures the similarity between two sets by dividing the size of their intersection by the size of their union, we choose the code snippet that retrieves the top ten scores each time to return as the prompt for the LLMs.

\subsection{Evaluation}

\begin{table}[t]
\centering \small
\caption{Experimental Results of Mitigation Method under Pass@1.}
\label{table:Mitigation-Experiment}
\tabmargin
\begin{tabular}{lrr}
\toprule
\textbf{Model} & \textbf{Raw Method} & \textbf{RAG-based Mitigation} \\ \midrule 
{CodeGen}        & 1.30\% & \cellcolor{blue!5}2.61\% ($\uparrow$ 1.31\%)\\
{PanGu-$\alpha$} & 0.04\% & \cellcolor{blue!5}1.74\% ($\uparrow$ 1.70\%)\\
{DeepSeekCoder}  & 3.04\% & \cellcolor{blue!5}3.91\% ($\uparrow$ 0.87\%)\\
{CodeLlama}      & 2.17\% & \cellcolor{blue!5}5.22\% ($\uparrow$ 3.05\%)\\
{StarCoder2}     & 0.04\% & \cellcolor{blue!5}2.61\% ($\uparrow$ 2.57\%)\\
{ChatGPT}        & 10.40\%& \cellcolor{blue!5}12.61\% ($\uparrow$ 2.21\%)\\
\bottomrule
\end{tabular}
\end{table}

We evaluate the effectiveness of the RAG-based mitigation method with the six LLMs: CodeGen, PanGu-$ \alpha $, ChatGPT, DeepSeekCoder, CodeLlama, and StarCoder2 on the CodeEval dataset. 
We compared our RAG-based mitigation method with the Raw method.
In the Raw method, we only provide LLMs basic docstrings and function signatures. In the RAG-based mitigation, when providing docstrings and function signatures, we will obtain ten related code snippets from the above-constructed retrieval library through a similarity algorithm as prompts and provide them to LLMs. We use the Pass@1 metric to assess the functionality correctness of the generated code snippets according to test cases. As shown in Table~\ref{table:Mitigation-Experiment}, the Pass@1 scores of all six models are consistently improved with the RAG-based mitigation method. 
Note that the performance improvement in our experiments is modest, as the mitigation method we explored is preliminary. We consider this experiment as an pilot study to explore the potential effectiveness of RAG-based mitigation. In future work, there are more methods worth studying, such as model fine-tuning and multi-agent framework with tool using, etc.

\begin{figure}[t]
    \centering
    % \captionsetup{labelfont=bf, name=Example, labelsep=period}
    \setlength{\abovecaptionskip}{0.1cm}
    \includegraphics[width=\linewidth]{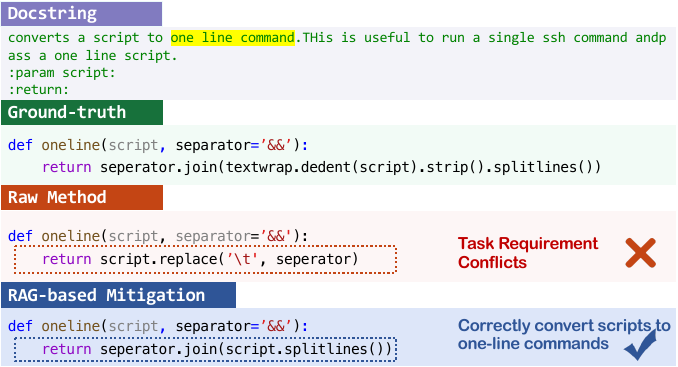}
    \figmargin
    \figmargin
    % \vspace{-1mm}
    \caption{Example: hallucination mitigation in Task Requirement Conflicts} %under the RAG approach}
    \figmargin
    \label{fig:method-compare-task-confilct}
\end{figure}

\begin{figure}[t]
    \centering
    % \captionsetup{labelfont=bf, name=Example, labelsep=period}
    \setlength{\abovecaptionskip}{0.1cm}
    \includegraphics[width=\linewidth]{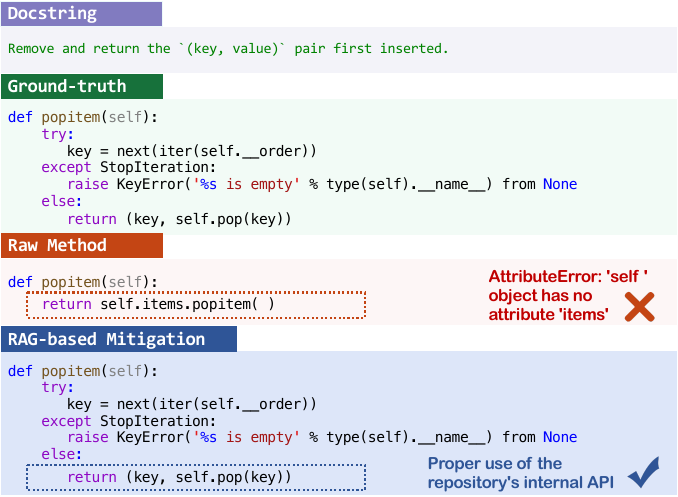}
    \figmargin
    \figmargin
    % \vspace{-1mm}
    \caption{Example: hallucination mitigation in Project Context Conflicts}% under the RAG approach}
    \figmargin
    % \Description{A PDF image showing Intention-Functionality Mismatch.}
    \label{fig:method-compare-api-confilct}
\end{figure}

To further illustrate the effectiveness of the hallucination mitigation, we conduct two case studies.
As shown in Figure~\ref{fig:method-compare-task-confilct}, in the Raw method, which only provides a docstring and a function signature, CodeGen incorrectly uses the \texttt{replace} function and fails to convert scripts to one-line commands. 
In contrast, with the RAG-based method, CodeGen correctly uses the \texttt{splitlines} function, aligning with the ground-truth and successfully addressing the requirement. In addition, the RAG-based method can also effectively mitigate Project Context Conflicts. As shown in Figure~\ref{fig:method-compare-api-confilct}, in the Raw method, ChatGPT attempts to use the \texttt{self.items.popitem()} API, which does not exist in the repository, leading to hallucinated generation. In contrast, with the RAG-based mitigation, ChatGPT correctly implements the requirement using the \texttt{self.pop()} function.

\section{Discussion}

We provide implications for future research on the hallucinations in practical LLM-based code generation.

\textbf{Developing hallucination identification techniques}: Through our study, we find 3 major %and eight minor 
categories of hallucinations in the LLM-based code generation. Some hallucinations can be detected by using static analysis (e.g., undefined variables) or dynamic test execution (runtime errors or test failures), making it relatively easy for developers to recognize and locate the relevant code issues. However, certain hallucinations, such as incomplete functionality and security issues, are very difficult for developers to detect and correct, as they can likely pass static checks and all test cases. As a result, LLM-generated code containing these hallucinations may be introduced into development projects and even real production environments, leading to unreliable software systems and severe security risks. Existing hallucination localization approaches~\cite{manakul2023selfcheckgpt, agrawal2023language} based on LLM self-feedback methods can detect hallucinations to a certain extent. However, these approaches heavily rely on the current model's capabilities and cannot address the fundamental limitations imposed by the training corpora. Therefore, in future work, researchers may consider developing more effective techniques to quickly and precisely identify and localize hallucinations in LLM-generated code.

\textbf{Developing more effective hallucination mitigation techniques}:
In Section~\ref{sec:mitgation_approach}, we explore the feasibility of applying a lightweight RAG-based method to mitigate hallucinations in LLM-based code generation. While the method demonstrates effectiveness in mitigating hallucinations such as undefined attributes, the potentials of RAG need to be further explored. For example, we only construct retrieval corpus using current code repository, leading to the augmented information is insufficient to mitigate many hallucinations such as background knowledge conflicts. In the future, we can integrate more comprehensive knowledge sources like online search engines, API documents, and StackOverflow discussions. In addition to RAG techniques, other methods such as input query refinement~\cite{dhole2023interactive, mu2023clarifygpt} and multi-agent systems~\cite{guo2024large} can also be leveraged to achieve an iterative process of (i) clarifying task requirements, (ii) generating code, (iii) running test cases, and (iv) mitigating hallucinations. To achieve this, we need to design the appropriate interaction protocols between agents and relevant tools (e.g., search engines and static analysis tools) and apply suitable prompting strategies.

\section{Threats to Validity}

\textbf{External Validity.} Threats to external validity mainly concern the generalizability of our findings. We focused on Python when exploring the taxonomy and root causes of hallucinations in LLM-based code generation due to its simplicity and ease of use. Constructing hallucination taxonomies for other programming languages and comparing them with our current taxonomy is a valuable future direction. Another potential threat is the limited scale of the adopted CoderEval dataset, which contains only 230 coding tasks. To mitigate this, we selected six LLMs and had each generate 10 code snippets for each task to ensure a sufficient number of annotations.

\textbf{Internal Validity.} Threats to internal validity primarily concern the manual annotation process in taxonomy construction. A key issue is the absence of formal inter-rater reliability measure for annotating hallucinations. To address this, discrepancies were discussed and resolved in annotator meetings to ensure a consistent annotation protocol, with each identified hallucination receiving a mutually agreed-upon label. Additionally, to ensure consistency in our findings, one author reviewed all labeled data. Another potential threat is model bias during the annotation process. To mitigate this, we mixed the generation results of the six models before annotation.

\textbf{Construct Validity.} Threats to construct validity are related to evaluating our hallucination mitigation approach. To alleviate these threats, we conducted experiments on six models using test cases available in the CoderEval dataset, a standard method for evaluating the correctness of generated code.

\section{Conclusion}
In this paper, we conduct an empirical study on code-generated hallucinations of large models in the practical development scenarios and through a full manual analysis, we construct a taxonomy of hallucinations and follow up with further hallucination classifications. Based on the hallucinations found, we provide a deeper discussion of the causes of hallucinations and the distribution of hallucinations in different LLMs. At last, we implement a RAG-based approach for hallucination mitigation and further discuss potential hallucination mitigation approaches.

% \begin{acks}
% This work is partially supported by fundings from the National Key R\&D Program of China (2022YFB2702203), the National Natural Science Foundation of China (No. 62032025), and Technology Program of Guangzhou, China (No. 202103050004).
% \end{acks}

\normalem
% \clearpage

% \bibliographystyle{ieeetr}
% \vspace{-0.2cm}
\bibliographystyle{IEEEtran}
\bibliography{ref}
\balance
% \bibliographystyle{ACM-Reference-Format}


\end{CJK}
\end{document}